\documentclass{article}

\usepackage{setspace}
\usepackage{PRIMEarxiv}
\usepackage{indentfirst}
\usepackage[utf8]{inputenc} 
\usepackage[T1]{fontenc}    
\usepackage{hyperref}       
\usepackage{url}            
\usepackage{booktabs}       
\usepackage{amsfonts}       
\usepackage{nicefrac}       
\usepackage{microtype}      
\usepackage{lipsum}
\usepackage{fancyhdr}       
\usepackage{graphicx}       
\graphicspath{{media/}}     
\usepackage{hyperref}
\usepackage{cleveref}
\usepackage[font=normalsize]{subfig}
\usepackage{stfloats}
\usepackage{multirow}
\usepackage{threeparttable}

\crefname{equation}{Eq.}{Eq.}
\crefname{figure}{Fig.}{Fig.}
\crefname{table}{Tab.}{Tab.}
\crefname{section}{Sec.}{Sec.}
\newcommand{\etal}{\textit{et al.}}

\usepackage{array}
\newcommand{\PreserveBackslash}[1]{\let\temp=\\#1\let\\=\temp}
\newcolumntype{C}[1]{>{\PreserveBackslash\centering}p{#1}}
\newcolumntype{R}[1]{>{\PreserveBackslash\raggedleft}p{#1}}
\newcolumntype{L}[1]{>{\PreserveBackslash\raggedright}p{#1}}

\title{Influence of Pre-bottleneck Diversion Devices on Pedestrian Flow
}

\author{
  Lu Wang\footnotemark[1],~~Saizhe Ding\footnotemark[1],~~Hang Yu,~~Xudong Li,~~Jun Zhang,~~Weiguo Song\footnotemark[2] \\
  University of Science and Technology of China \\
  Hefei, Anhui, 230026, P.R.China. \\
}

\begin{document}
\maketitle

\renewcommand{\thefootnote}{\fnsymbol{footnote}}
\footnotetext[1]{These authors contribute equally to this work.}
\footnotetext[2]{Corresponding author (e-mail: wgsong@ustc.edu.cn).}

\begin{abstract}
The existence of bottlenecks often leads to the stagnation of pedestrian gatherings, which seriously affects the efficiency of traffic and reduces the flow of pedestrians. Some studies have shown that setting devices in front of bottlenecks can promote pedestrian evacuation under certain conditions. In this paper, the effect of setting diversion devices in front of the exit on pedestrian flow is studied. From our observation, these diversion devices can form a buffer zone before the exit and affect pedestrian behaviors. The evacuation times are found to decrease as the devices become farther away from the exit. In our experiments, it is found that the effect of shunt piles on evacuation is better than in the case of safety barriers and without device conditions. Under the condition of setting up safety barriers approximately $1~m$ and $3~m$ in front of the exit, the evacuation times are extended by $0.88\%$ and $2.67\%$. For shunt piles, the evacuation times are $11.53\%$ and $14.96\%$ shorter than that of those without a device regarding the different distances to exit ($1~m$ and $3~m$, respectively). In addition, setting up shunt piles reduces the time interval between two consecutive pedestrians. To sum up, in our experimental settings, the diversion devices can effectively improve the average speed ahead of the exit and promote evacuation to become more orderly, which reduces the congestion in the later period of evacuation. In other words, this study demonstrates that a reasonable layout of facilities can not only meet the daily functional requirements but also improve the efficient use of space in emergencies, reducing the probability of crowd conventions and jams.
\end{abstract}


\section{Introduction}
\label{sec:introduction}

Bottlenecks are construction facilities (e.g. exits) that limit the movement of people. When the crowd density is high, the traffic flow will be limited by the bottleneck structure, leading to the decline of evacuation efficiency, and further bringing congestion and even stampede accidents. In order to reduce the risk of accidents at the bottleneck, many scholars \cite{helbing2000simulating,kirchner2003friction,shiwakoti2019review} have conducted many experiments and simulations related to the bottleneck. Generally, many kinds of research mainly focus on the impact of evacuation by changing the types of bottlenecks. For example, in \cite{hoogendoorn2005pedestrian, kretz2006experimental, seyfried2009new, liddle2009experimental, song2013experiment, liao2014experimental, sun2017comparative, tavana2019insights}, they study how the width, length, and other parameters of bottleneck influence evacuation efficiency, which find that some self-organizing phenomena, such as lane formation, zipper effect, faster-is-slower effect and so on, are highly correlated to the above factors. Besides, under the condition of crowd competition, the frequency of congestion increases with the decrease in bottleneck width \cite{muller1981gestaltung}. When the bottleneck width is small, the competitive behavior will reduce the evacuation efficiency \cite{muir1996effects}. In order to further match the real situation, some researchers \cite{tsuchiya2007evacuation, shimada2006experimental, ren2019experimental, li2020comparative} add the extra factors of the proportion of different ages, genders, special people, etc.

On the other hand, some scholars consider setting a reasonable diversion device before the exit to relieve the competitive pressure of pedestrians at the bottleneck, so as to improve evacuation efficiency and reduce the risk of accidents. For example, Helbing \etal \cite{helbing2005self} study the effect of adding obstacles in front of exits on the evacuation effect. They design an experiment on the evacuation of pedestrians in a room with an $82~cm$ wide door and place a wooden board with a width of $45~cm$ as an obstacle near the exit. The results show that the obstacle increased the flow of pedestrians in the room by approximately $30\%$ compared to no obstacle, and the time interval between two consecutive pedestrians leaving the room become shorter. Haghani \etal \cite{haghani2019simulating} use a social force model to perform extensive simulations to find the optimal parameters of the model and fit the relationship between pedestrian evacuation speed and time to derive quantitative results between the two. Echeverría and Zuriguel \cite{echeverria2020pedestrian} perform numerical simulations of an asymmetric spherical column system. When the distance between the obstacle and the wall is smaller than the width of the exit, the evacuation time is prolonged; by increasing the distance from the obstacle to the exit, the blockage can be reduced, and the optimal obstacle location can be obtained. Daichi Yanagisawa \etal [19] study the effect of obstacles in front of the exit in NHK television studio in Japan. A $20~cm$ diameter post is placed in front of the $50~cm$ wide outlet. It is found that participants have a greater outflow (approximately $7\%$ increment) when the barrier is in front of the exit compared to normal evacuation. Wang  \etal [20] find that setting railings in front of the exit to form a buffer zone can improve evacuation efficiency, effectively reduce the crowd density at the exit, alleviate the inrush of density waves in the crowd, and thus reduce the risk of crowd extrusion and stampede. In addition, when the desired speed is high, the longer the buffer, the faster the pedestrian evacuation speed. Jiang \etal [36] conducted experiments on 76 college students (aged 21-25 years) to study the influence of obstacles near exits. One obstacle, two obstacles, and no obstacle are set before the exit of the experiment. It turns out that there are two barriers near the exit that work best. In 2020, Zhao \etal \cite{zhao2017optimal} simulate the effects of different shapes of obstacles through a social force model, analyse the density map, velocity map, flow map, and spatiotemporal pressure map of placing obstacles at the exit, and conclude that placing obstacles at the placement of obstacles at the exit can effectively reduce the crowd density, reduce the blockage effect at the exit bottleneck, and improve the evacuation efficiency. The robustness and stability of plate-shaped obstacles are better than column-shaped obstacles in complex scenarios, while the crowd evacuation efficiency is found to be greatly influenced by the geometric parameters of the obstacles.

However, not all findings suggest that the placement of barriers before exits has a positive effect. Garcimartín \etal \cite{garcimartin2015flow, garcimartin2018redefining}. vary the degree of emergency (as reflected by different desired pedestrian speeds) to study the effect of barriers. The results show that the effect of barriers is negative in low as well as high competing pressures, inhibiting pedestrian evacuation and prolonging the overall time. Varas \etal \cite{varas2007cellular}. conduct a simulation experiment on pedestrian evacuation under barriers using a 
Cellular Automata model and obtain the same results, where the presence of barriers adversely affects pedestrians. Liu's \cite{liu2016controlled} experiment shows  that in the case of congestion, placing cylindrical barriers in front of safety exits will eliminate the clustering phenomenon of pedestrians, but reduce the evacuation efficiency. Chattaraj \etal \cite{chattaraj2013empirical} investigate the effect of the presence of strip space barriers in pedestrian walkways on the local speed and density of pedestrians and conclude that the placement of strip barriers along the walking direction of pedestrians in the walkway will reduce the evacuation efficiency of pedestrian flow. In summary, the effect of obstacles on evacuation efficiency is related to their size, location, material, and experimental conditions, and the experimental results may differ in different cases.

At present, the most commonly used pedestrian dispersal measures in public spaces with large numbers of people and crowded areas and their internal "bottleneck" areas are to change the distribution of pedestrians in the passage. To maintain a safe flow of passenger traffic by controlling pedestrian routes at appropriate times \cite{yang2016safety}. The most commonly used methods are batch release, flow control at entrances and exits, channel space flow separation, etc. Adding barriers such as separation rails in front of channel exits, usually by setting flow restriction rails to slow down passenger flow, can effectively improve the orderliness and stability of passenger queues \cite{xu2014analysis} and facilitate on-site staff to maintain order.

However, there is a lack of research on whether safety barriers are the best choice, in which position they have the best evacuation effect, and whether they will have a negative impact when an emergency occurs. Therefore, this paper carried out a series of pedestrian evacuation experiments. By setting different types of diversion devices in the experimental scene and changing the distance between them and the exit, so as to explore the influence of the diversion devices on the pedestrian movement law.

\section{Experiment Setup}
\label{sec:experiment}

To investigate the effect of the above diversion devices on pedestrian evacuation efficiency, we design an experimental scenario, as shown in \cref{tab:category}(a). The experimental scenario is a closed structure with a one-way through lane, the pedestrian waiting area is delineated on the left side of the scenario, and the bottleneck is set on the opposite side of the waiting area with a width of $0.6~m$, which meets the minimum width of single passenger flow at the safety exit. Before the start of the experiment, participants are required to stay in a row in the waiting area, and after hearing the instructions, all participants have to quickly move through the closed area towards the bottleneck and leave the experimental area.

\begin{figure}[htbp]
    \centering
    \subfloat[]{\includegraphics[width=0.51\linewidth]{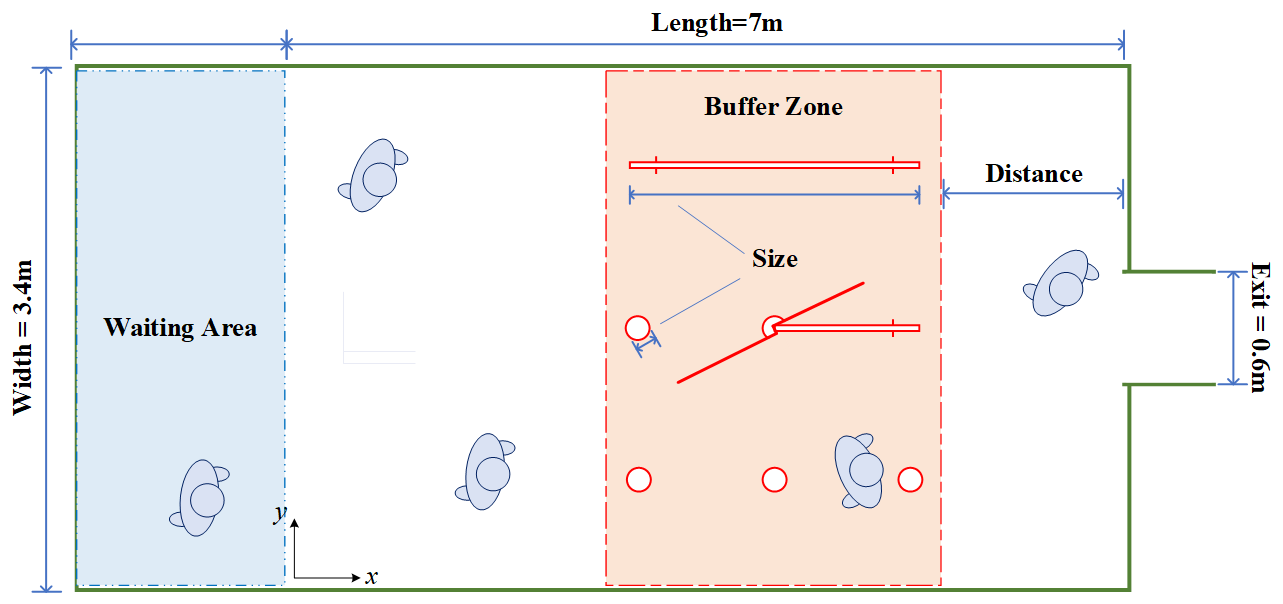}}\hfill
    \subfloat[]{\includegraphics[width=0.49\linewidth]{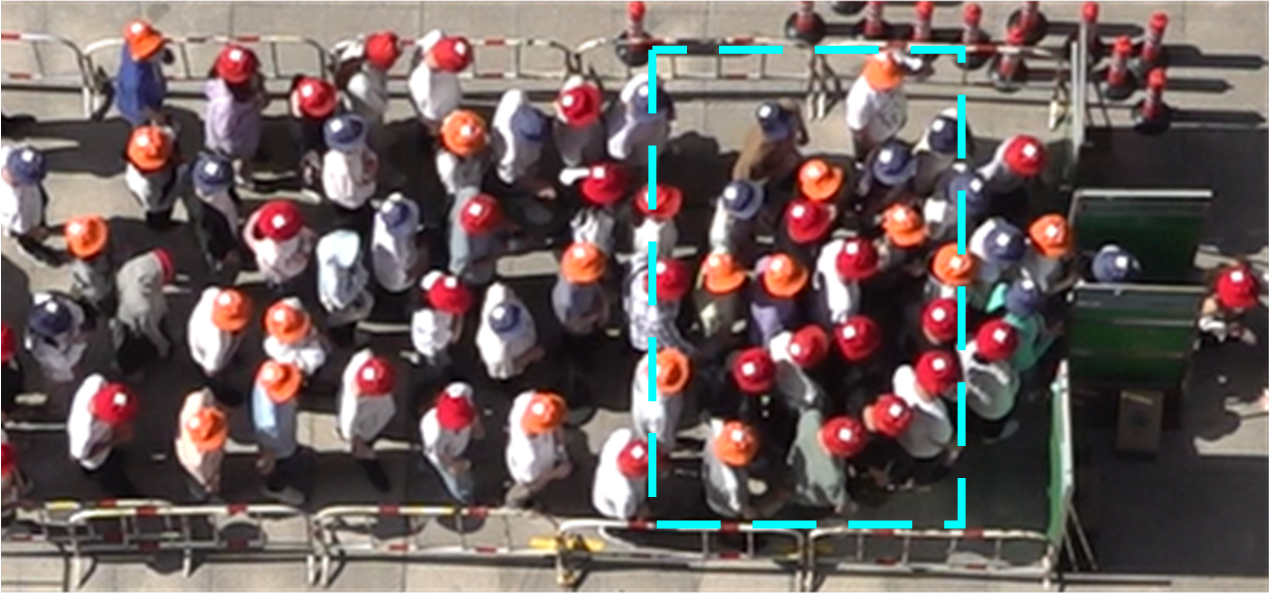}}
    \caption{Video screenshot of the experimental scenario and mapping of the scenario. Eighty-seven adults (21-25 years old) initially stood in the waiting area with the device placed $x$ meters in front of the exit. The buffer zone is the orange area where the device is placed in front of the bottleneck.}
    \label{fig:setup}
\end{figure}

To simulate the possible scene distribution, in reality, we chose to use six shunt piles as well as safety barriers as diversion devices (D) and set them at $1~m$ and $3~m$ from the bottleneck, where the radius of the shunt pile is $0.1~m$ and the safety barrier is $1.50~m$ long, both of which are conventional sizes. Through several experiments, the device type, as well as the distance to the bottleneck, were changed to obtain experimental data under different working conditions, and all the experimental conditions are shown in \cref{tab:category}. The experiment was conducted in September 2021 in Hefei, Anhui Province, China, as shown in \cref{fig:setup}(b). Eighty-seven student volunteers with an average age of $23 \pm 3~years$ $(20\sim26~years)$ and an average height of $170.2~cm$ $(155\sim185~cm)$ joined the experiment. Each volunteer was asked to wear a colored hat (red, orange, and blue) during the experiment, which was recorded by two digital cameras (model: HDR-SR11, resolution: $1920 \times 1080 ~pixels$, frame rate: $25~fps$) set up at a $30~m$ height from the ground. The captured video was lens-deformed through a standard chessboard grid $5 \times 7 ~(50~mm)$, followed by automatic extraction of pedestrian trajectory data by PeTrack \cite{boltes2010automatic} software. and the average pedestrian height of $170.2~cm$ was used as a reference for data conversion from pixel coordinates to physical coordinates during the extraction process.

\begin{table}[htbp]
\caption{Experimental working condition table}
\centering
\begin{threeparttable}
\begin{tabular}{r|C{2.5cm}C{2.5cm}C{3cm}C{2.5cm}}
\toprule
Index \tnote{*} & Device            & Size    & Distance from Exit    & Amount    \\
\midrule
B-1             & Safety   barriers & 1.5 $m$   & 1 $m$                   & 1.5 $m$     \\
B-3             & Safety   barriers & 1.5 $m$   & 3 $m$                   & 1.5 $m$     \\
P-1             & Shunt piles       & 0.1 $m$   & 1 $m$                   & 1.5 $m$     \\
P-3             & Shunt   piles     & 0.1 $m$   & 3 $m$                   & 1.5 $m$     \\
N-0             & No device         & -       & -                     & -         \\
\bottomrule
\end{tabular}
\begin{tablenotes}
\footnotesize
\item[*] where the safety barriers are indicated by B, the shunt piles are indicated by P, and no device is indicated by N; the control devices in this thesis are of the same length of 1.5 $m$, and the number after the letter represents the distance from the device to the exit.
\end{tablenotes}
\label{tab:category}
\end{threeparttable}
\end{table}

\section{Spatial Distribution Characteristics}
\label{sec:results}

In this chapter, we analyze the above-mentioned experimental data by using JPSreport \cite{wagoum2015jupedsim} and Python processing. In \cref{subsec:evac_time}, we firstly conduct an overall cross-sectional comparison of all working conditions to analyze the effect of different working conditions on pedestrian evacuation time. To further investigate the pattern, we plot the pedestrian trajectories for different working conditions in \cref{subsec:ped_traj} to observe the effect of the device on the pedestrian movement patterns.

\subsection{Time Analysis}
\subsubsection{Evacuation Time}
\label{subsec:evac_time}

In \cref{tab:evacuation}, we list the average evacuation time for each condition. To ensure the fairness of the comparison, the average evacuation time is taken as the average of three repeated experiments. To eliminate errors, the first 5 people and the last 5 people in each run are removed to avoid the transient state. As seen in \cref{tab:evacuation}, the evacuation time always decreases with increasing distance between the diversion devices and the exit, regardless of the type of diversion devices used. When we consider the type of diversion devices, it can be found that the evacuation times of safety barriers are all longer than that of no device, increasing by $1.28~s$ and $0.42~s$, respectively (evacuation efficiency decreases by $2.67\%$ and $0.88\%$), while shunt piles facilitate evacuation, reducing the overall evacuation time by $5.52~s$ and $7.16~s$, respectively (evacuation efficiency increases by $11.53\%$ and $14.96\%$). Therefore, we believe that when the distance between the diversion devices and the exit is short, the use of safety barriers as a diversion device is not recommended.

\begin{table}[htbp]
\caption{Overall evacuation time of pedestrians under different working conditions}
\centering
\begin{threeparttable}
\renewcommand{\arraystretch}{1.3}
\begin{tabular}{r|C{2cm}C{2cm}cC{2cm}c}
\toprule
Distance to exit & $0~m$                        & $1~m$     & Improvement\tnote{*}   & $3~m$     & Improvement\tnote{*} \\
\midrule
Safety barriers  & \multirow{2}{*}{$47.88~s$}   & $49.16~s$ & $-2.67\%$              & $48.30~s$ & $-0.88\%$            \\
Shunt piles      &                              & $42.36~s$ & $+11.53\%$             & $40.72~s$ & $+14.96\%$           \\
\bottomrule
\end{tabular}
\begin{tablenotes}
\footnotesize
\item[*] In "Improvement", "+" means that the evacuation time is improved compared with no device; “-" represents reduced evacuation time compared to no device efficiency.
\end{tablenotes}
\label{tab:evacuation}
\end{threeparttable}
\end{table}

\subsubsection{Time Interval}
\label{subsec:interval}

\begin{figure}[htbp]
\centering
\includegraphics[width=0.5\linewidth]{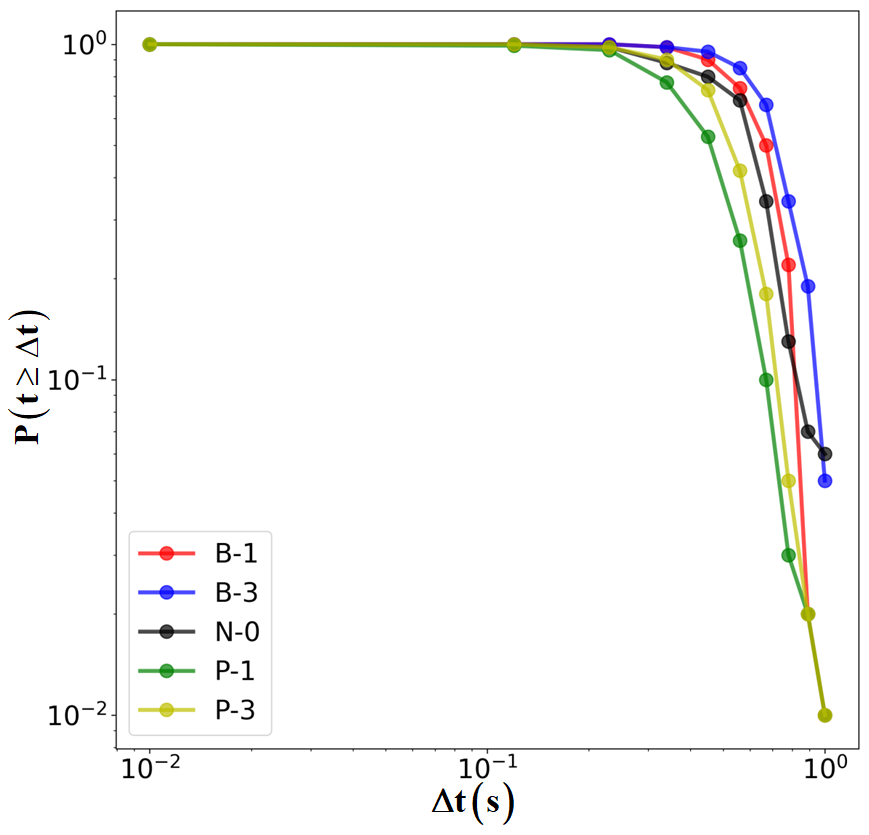}
\caption{Cumulative probability of the time interval $\Delta t$.}
\label{fig:time_interval}
\end{figure}

The former studies in \cite{clauset2009power, shi2019examining} show that the shorter the time interval between two consecutive pedestrians leaving the exit is, the smoother the pedestrian leaves the bottleneck. Referring to \cite{clauset2009power}, we study the variation of the time interval $\Delta t$ based on the Clauset-Shalizi-Newman method, and the formula is as follows:

\begin{equation}
\Delta t = t_{i} - t_{i - 1}
\label{eqn:time_interval}
\end{equation}

where the time interval $\Delta t$ is defined as the time elapsed between two consecutive pedestrians passing through the bottleneck. To eliminate the experimental error, we delete the data of the first five pedestrians and the last five pedestrians in each experiment because they all represent transients and have large randomness. In general, the $p(\Delta t)$ distribution conforms to the power-law distribution formula $p(\Delta t) \sim \Delta t ^ {- \alpha}$, while the exponent $\alpha$ can be used to measure the dynamics of the exit \cite{pastor2015experimental} and the slope of the curve is related to the congestion level. The higher the slope is, the longer the tail of the curve, which means the longer the congestion time. Therefore, we draw the cumulative probability map of different working conditions, as shown in \cref{fig:time_interval}.

\begin{table}[htbp]
\caption{The exponent $\alpha$ obtained from the power-law analysis with $x_{min}=0.5$}
\centering
\renewcommand{\arraystretch}{1.3}
\begin{tabular}{C{3cm}|C{3cm}C{3cm}}
\toprule
Index & $\alpha$ & $\Delta t$ \\
\midrule
N-0   & 2.31     & 0.58      \\
B-1   & 2.27     & 0.55      \\
B-3   & 2.32     & 0.57      \\
P-1   & 2.37     & 0.51      \\
P-3   & 2.43     & 0.48      \\
\bottomrule
\end{tabular}
\label{tab:exponent}
\end{table}

See \cref{tab:exponent}. It implies that the $\alpha$ larger value at a narrow exit indicates a smaller probability of clogs. The increase of the $\alpha$ value under the working conditions of the shunt piles indicates that a reasonable shunt device can promote the continuity of pedestrian outflow at the exit. And the time interval between two consecutive pedestrians passing through the exit is significantly shortened, so that the flow of people can pass through the exit more efficiently.

\subsection{Pedestrian Trajectories}
\label{subsec:ped_traj}

\begin{figure}[htbp]
\centering
\subfloat[]{\includegraphics[width=0.5\linewidth]{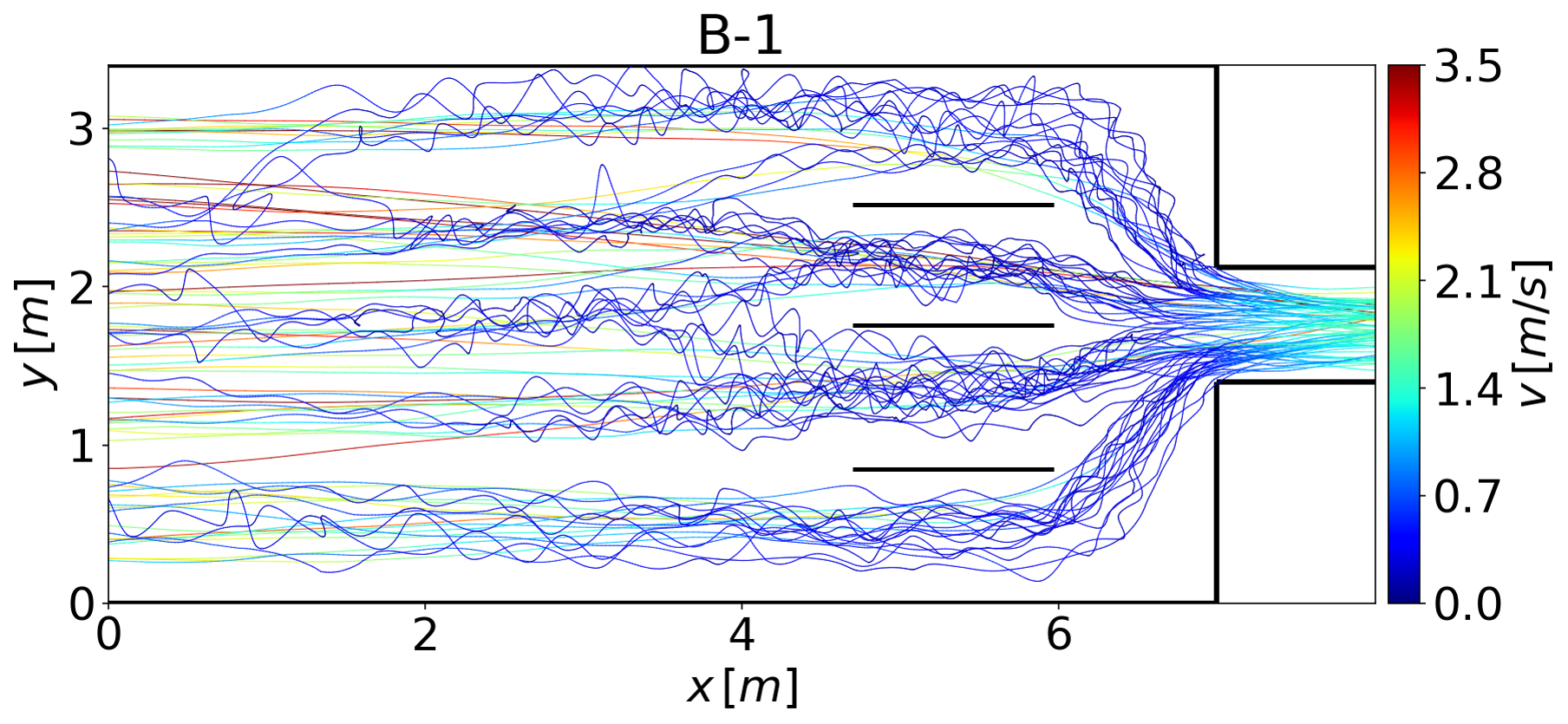}}\hfill
\subfloat[]{\includegraphics[width=0.5\linewidth]{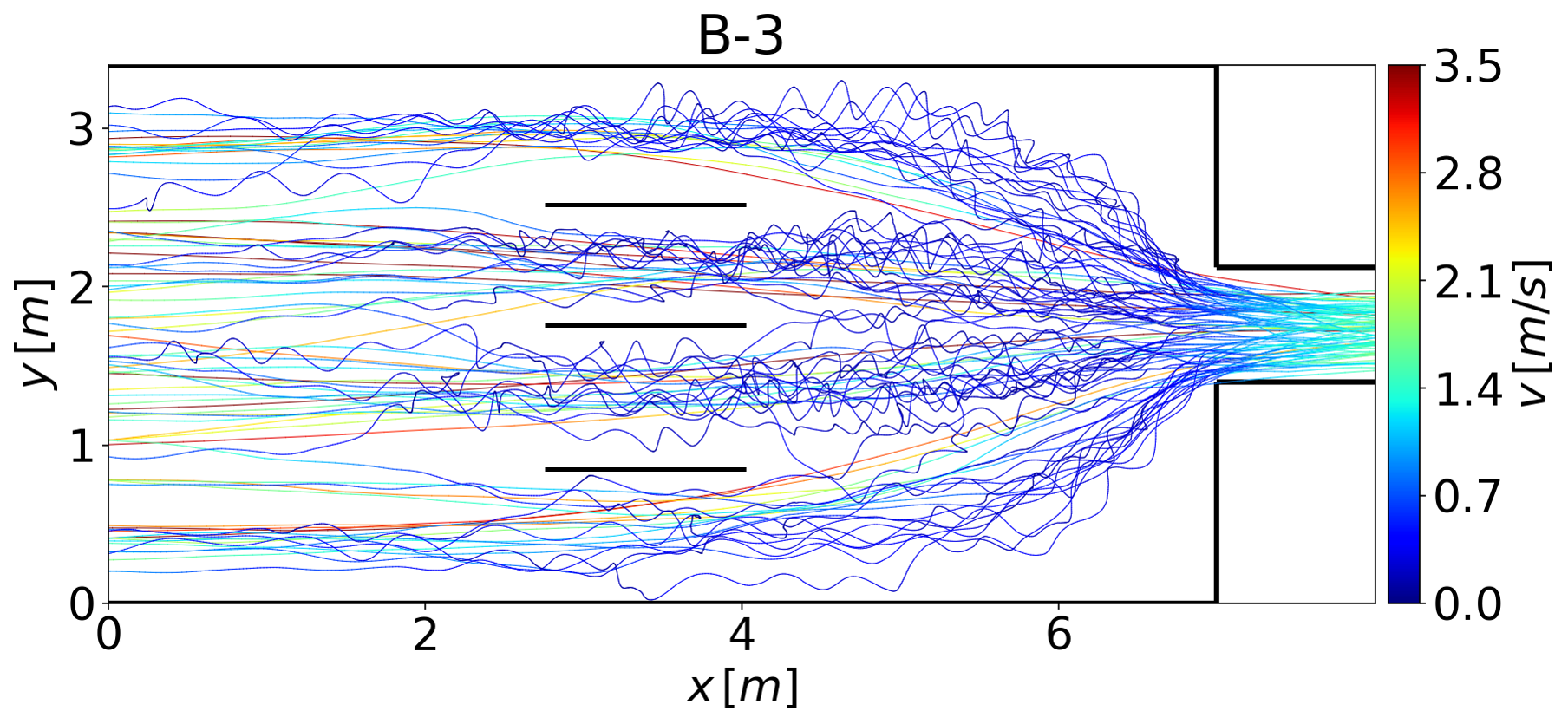}}\\
\subfloat[]{\includegraphics[width=0.5\linewidth]{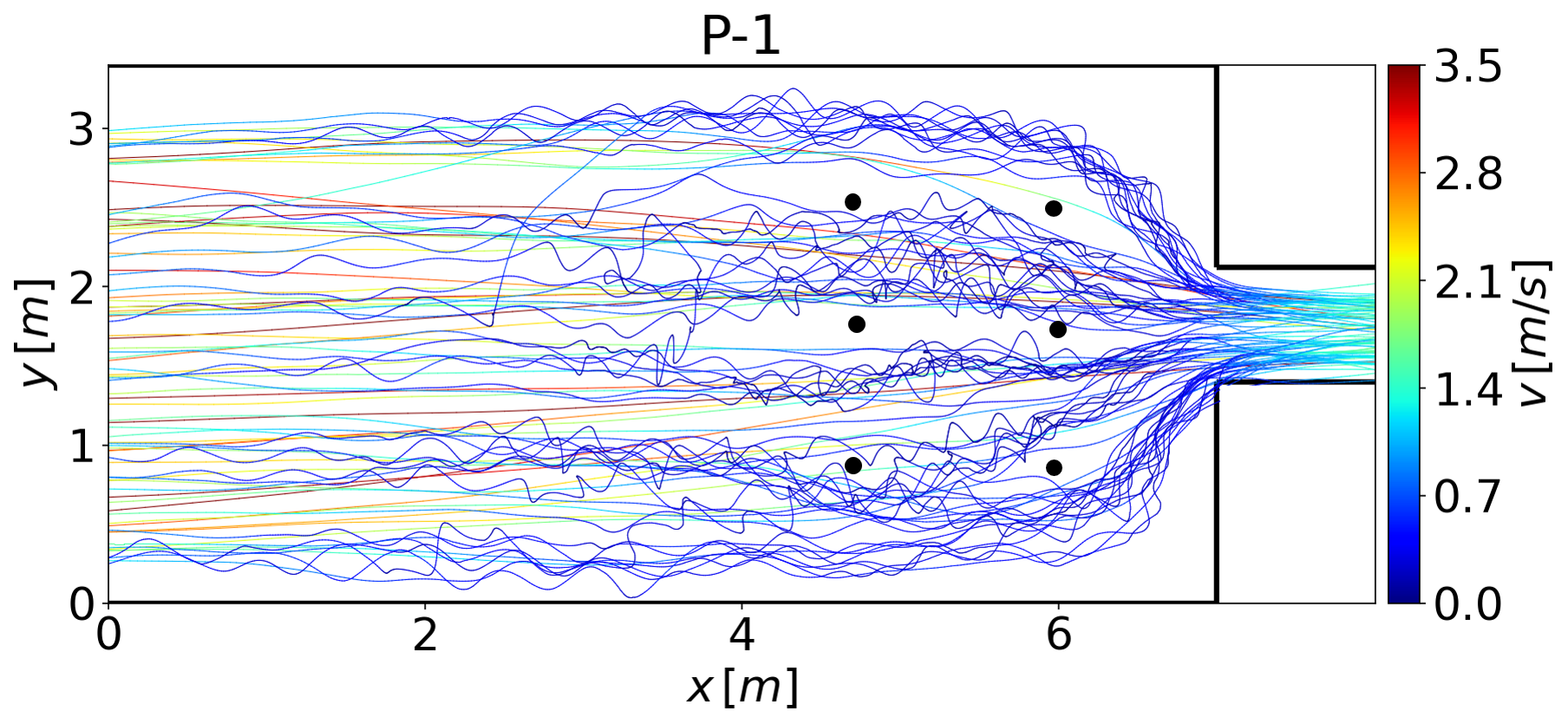}}\hfill
\subfloat[]{\includegraphics[width=0.5\linewidth]{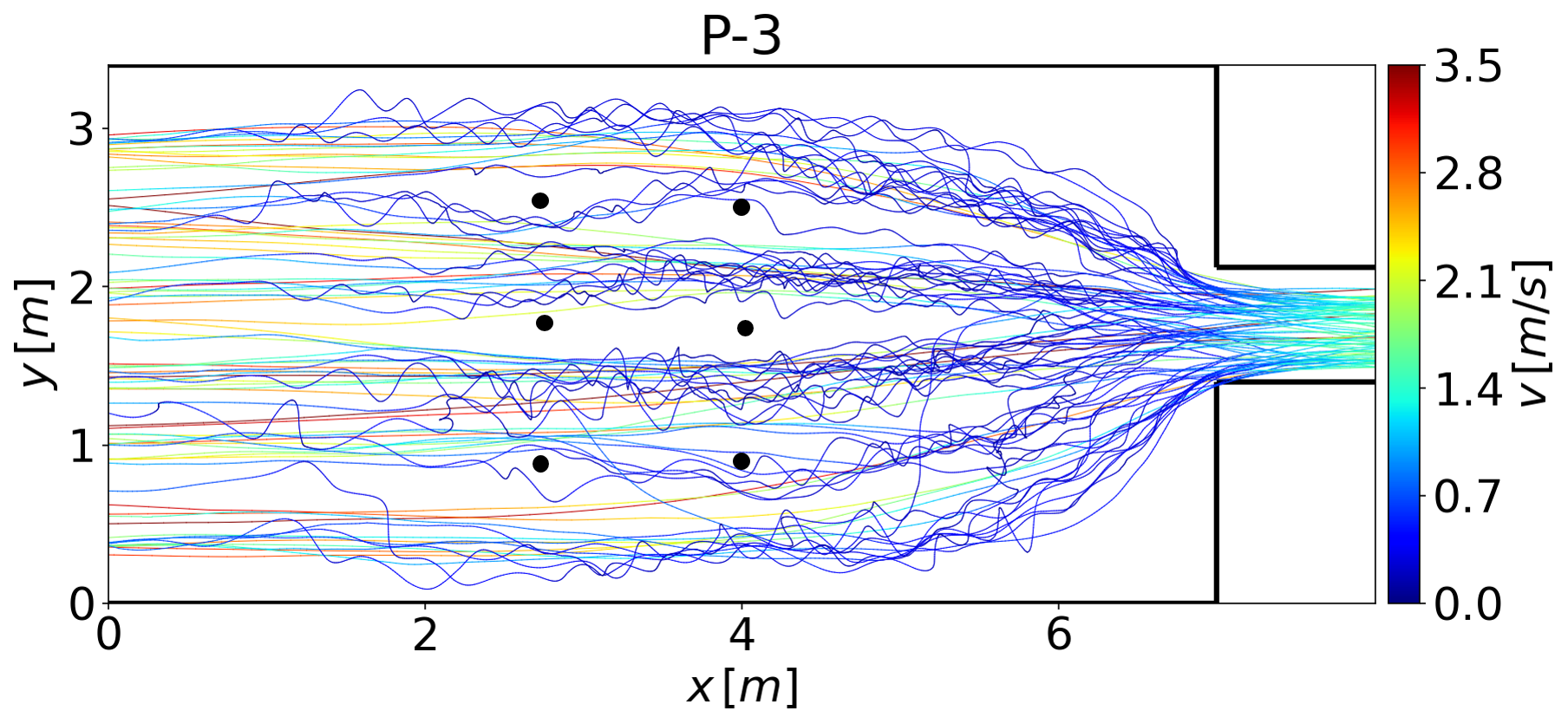}}\\
\subfloat[]{\includegraphics[width=0.5\linewidth]{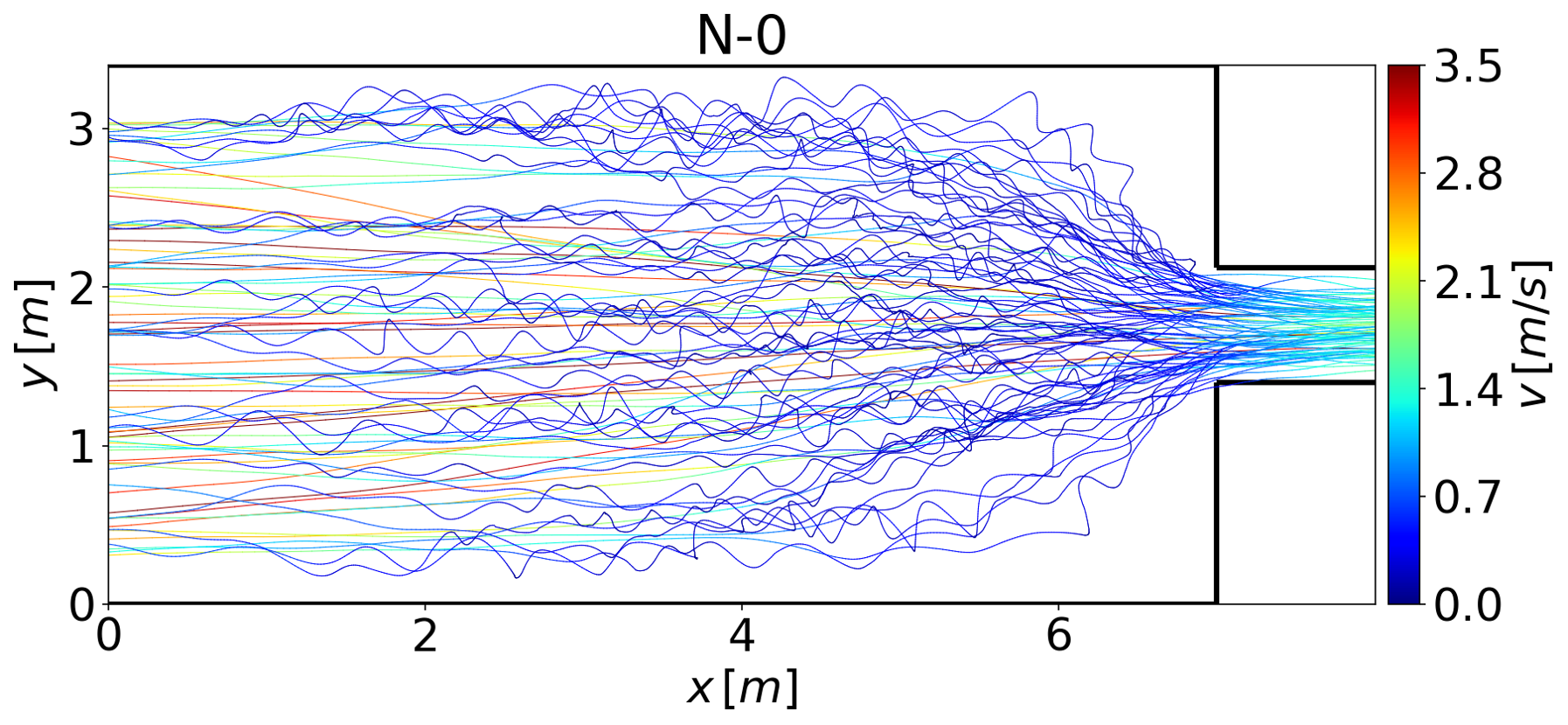}}
\caption{Trajectory diagram of pedestrians under different working conditions.}
\label{fig:trajectories}
\end{figure}

To investigate the intrinsic reasons for the influence of device type and placement distance on pedestrian movement characteristics, we plotted the pedestrian trajectories in different scenarios, as shown in \cref{fig:trajectories}. Different colors represent the different speeds of movement of pedestrians. Experimental video can be found that almost all scenarios follow the phenomenon that when the evacuation starts, people in the front row of the queue move rapidly toward the exit, and since the population density is still relatively low at this time, they can proceed almost unimpeded. After a few seconds, as the density near the exit increases, congestion begins to form in front of the exit. By comparing the shape of the trajectory, we find that there are differences in whether the shunt piles or safety barriers have an impact on the movement trajectory of pedestrians, as shown by the different positions of pedestrians contracting toward the exit.

\subsubsection{Trajectory Deflection}

With the gradual increase of the distance to the exit for the device, the possibility of physical contact among pedestrians and the walls on both sides of the exit decreases. The trajectory profile formed at the exit gradually shrinks, slowly showing the same teardrop type distribution as without the device (such as \cref{fig:trajectories} (b) and \cref{fig:trajectories} (d)), but although these several working conditions (\cref{fig:trajectories} (b) and (d)) have the same shape distribution at the exit, for the shunt piles and safety barriers in the evacuation efficiency there is a significant difference, this phenomenon can be explained by the phenomenon of pedestrians waiting in front of the exit, when the pedestrian speed is slow the body will appear to sway from side to side to maintain body balance, and the trajectory will appear to deflect from side to side.

\subsubsection{Differences in Lane Change Behavior}

Unlike the case without the device, under the condition of shunt devices, the pedestrian trajectories automatically form four lanes (before reaching the device placement area) from the state of dispersion (before reaching the device placement area, hereafter referred to as the buffer zone), thus creating a stratification phenomenon \cite{lee2016modeling, feliciani2016empirical}. Since shunt piles have greater flexibility compared to safety barriers, it is easier for pedestrians to change their routes under this device. Therefore, due to the speed impact of queuing, it can be eliminated by changing the path selection, so as to achieve more flexible space selection and improve evacuation efficiency. Lane changing behavior is more pronounced when set 1 $m$ before the exit, with 9 pedestrians changing their path choice before reaching the safety barriers, and in the case of the shunt piles, pedestrians mostly change their path inside the device area, with 9 people and only 3 people changing their path before reaching the shunt piles. The number of people changing lanes at the safety barriers set at $3~m$ was 2, and the number of people changing lanes at the shunt piles was 10.

\subsubsection{Differences in Space Utilization}

\begin{figure}[htbp]
\centering
\includegraphics[width=0.5\textwidth]{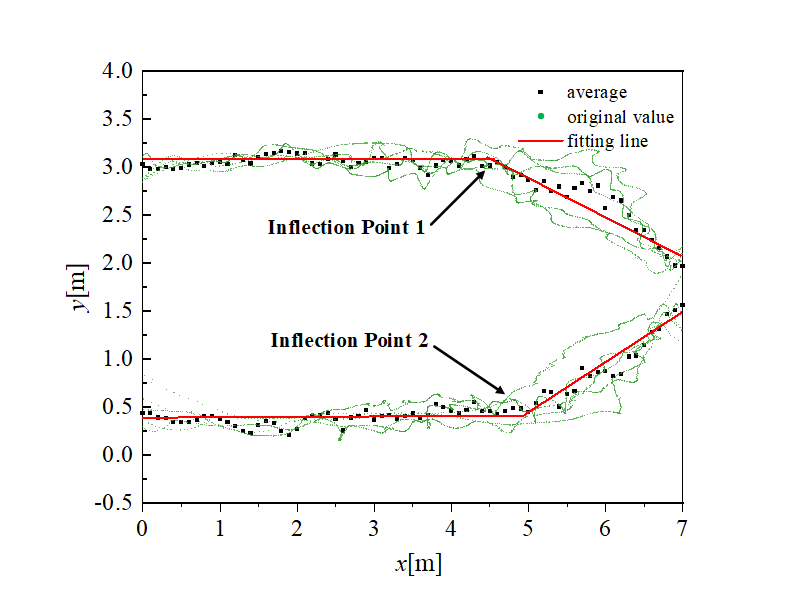}
\caption{Trajectory diagram of pedestrians under different working conditions.}
\label{fig:inflection}
\end{figure}

The diversion devices have an impact on the trajectory of the pedestrians on the outside of the channel, mainly affecting the timing of crowd contraction before the exit, which determines the difference in space utilization before the exit and is also a key factor affecting evacuation. Therefore, we extracted the movement trajectory of the outermost 10 pedestrians (5 on each side). The scatter plot of the average trajectory can be obtained by calculating the mean y under the same x coordinate, as shown in \cref{fig:inflection}. By using the piecewise linear fitting method, we can obtain the location of the inflection point where the pedestrian trajectory turns, as shown in \cref{tab:linear_fit}. As seen from the table, under the condition of shunting piles with the same device distance, the position of pedestrian contraction always occurs earlier than that of safety barriers. When the devices are $1~m$ away from the exit, the pedestrian starts to adjust the direction to approach the exit at about $20\sim30~cm$ from the end of the safety barriers (B-1), while it is approximately $50\sim60~cm$ in the case of shunt piles (P-1). This indicates that pedestrians may be more inclined to maintain a certain distance from the safety barriers arranged in the passage. From the evacuation time of \cref{subsec:evac_time}, it can be seen that the shorter the evacuation time, the earlier the inflection point of the pedestrian trajectory It may be that the sooner the outside pedestrians approached the exit, the less likely it was to create congestion at the exit, and our suspicions can be confirmed in the video. It can also be seen in the table that the device appears earlier at the inflection point at 3 $m$ (B-3, P-3) than at $1~m$ (B-1, P-1). and the evacuation time at $3~m$ is shorter than that at $1~m$.

\begin{table}[t]
\caption{Coordinates of the outermost pedestrian inflection point under different working conditions and linear fit correlation}
\centering
\begin{tabular}{R{0.05\textwidth}|C{0.2\textwidth}C{0.2\textwidth}C{0.2\textwidth}C{0.2\textwidth}}
\toprule
Index & Inflection Point 1    & $R^2$   & Inflection Point 2    & $R^2$ \\
\midrule
N-0   & (4.84, 0.37)          & 0.94    & (4.61, 3.05)          & 0.92  \\
B-1   & (5.82, 0.36)          & 0.95    & (5.72, 3.13)          & 0.93  \\
B-3   & (5.16, 0.38)          & 0.96    & (4.88, 3.06)          & 0.96  \\
P-1   & (5.49, 0.35)          & 0.85    & (5.35, 3.09)          & 0.86  \\
P-3   & (4.71, 0.35)          & 0.97    & (4.06, 3.10)          & 0.99  \\
\bottomrule
\end{tabular}
\label{tab:linear_fit}
\end{table}

\begin{figure}[htbp]
\centering
\subfloat[]{\includegraphics[width=0.5\linewidth]{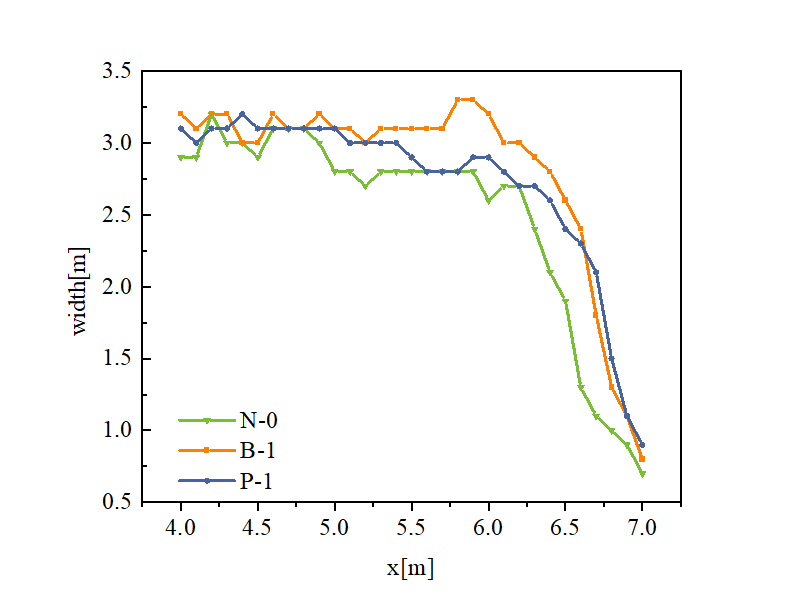}}\hfill
\subfloat[]{\includegraphics[width=0.5\linewidth]{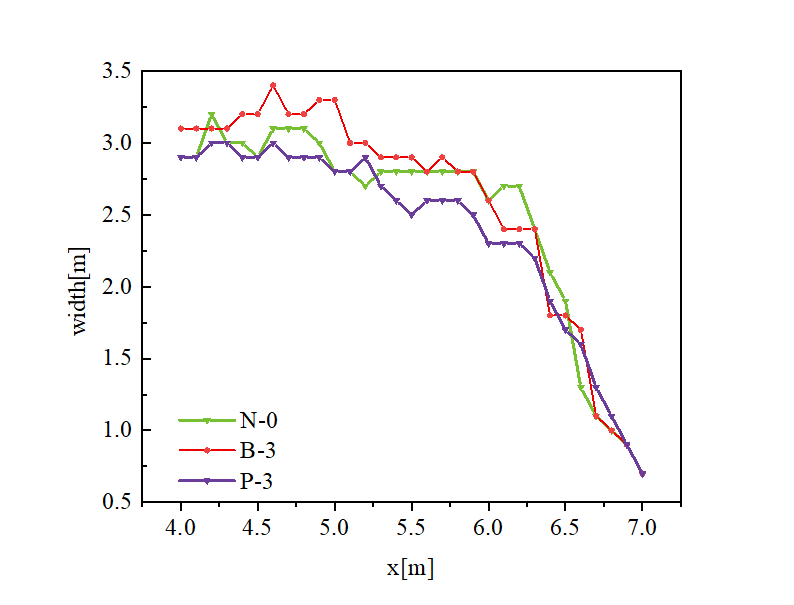}}\\
\subfloat[The difference in the width of the crowd]{\includegraphics[width=0.5\linewidth]{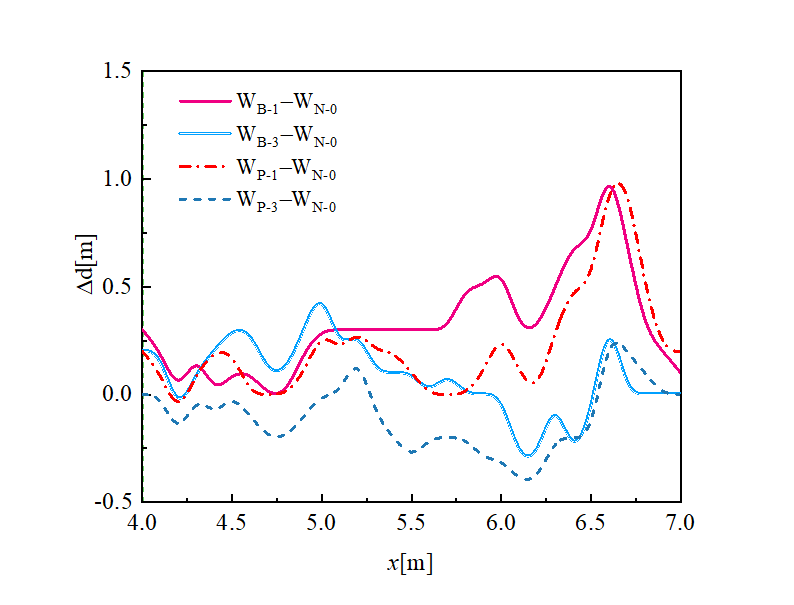}}
\caption{Trajectory diagram of pedestrians under different working conditions.}
\label{fig:width}
\end{figure}

We select the earliest point $x$ ($x$=4.06) that starts to shrink to study the behavior of the crowd after the team shrinking. It mainly analyses the changes in the width of the crowded queue, which is defined as the length of the crowded queue in the ordinate direction in the coordinates of the experimental scene. \cref{fig:width} records the variation in the width of the crowd in different scenarios (\cref{fig:width} (a) - (b)) and the difference between the width of the crowd line relative to the deviceless situation (\cref{fig:width} (c)). When the safety barriers (B-1) are placed $1~m$ in front of the exit, the width of the crowd in the buffer area is higher, especially before reaching the devices. Pedestrians begin to consciously increase the distance between each other to enter the buffer area, and the width gradually rises to a peak after leaving the buffer area. In contrast, the difference between the width of the crowd of the diversion safety barriers placed at $1~m$ and the width of the crowd without the device is not as significant as that of the shunt piles (P-1). When the shunting device is placed $3~m$ in front of the exit, the width of the crowd of the safety barriers (B-3) is still higher than that without the device for most of the time, while the shunt piles (P-3) show a shrinking phenomenon. According to the research of Garcimartín \etal \cite{garcimartin2015flow,garcimartin2018redefining}, the more orderly the speed and direction of the crowd are, the shorter the time interval of one-way flow through the exit, that is, the faster pedestrian flow. As seen from the trajectory (in \cref{fig:trajectories}.), when the diversion devices are far away from (with $3~m$) the exit, the pedestrians near the exit are easier to shrink and difficult to form an arch, which makes the speed direction of the crowd behind the exit tend to be consistent, so the evacuation efficiency can be improved. The analysis in \cref{subsec:interval} further validates our idea.

\section{Characteristic Analysis of Evacuation Motions}
\label{sec:analysis}

In this section, we first analyse the various characteristics of movement quantities in \cref{subsec:characteristics}, such as pedestrian density, in \cref{subsection:CL} to investigate the congestion level of pedestrian movement.

\subsection{Pedestrian Density and Motion Characteristics}
\label{subsec:characteristics}

\begin{figure}[htbp]
\centering
\includegraphics[width=0.5\linewidth]{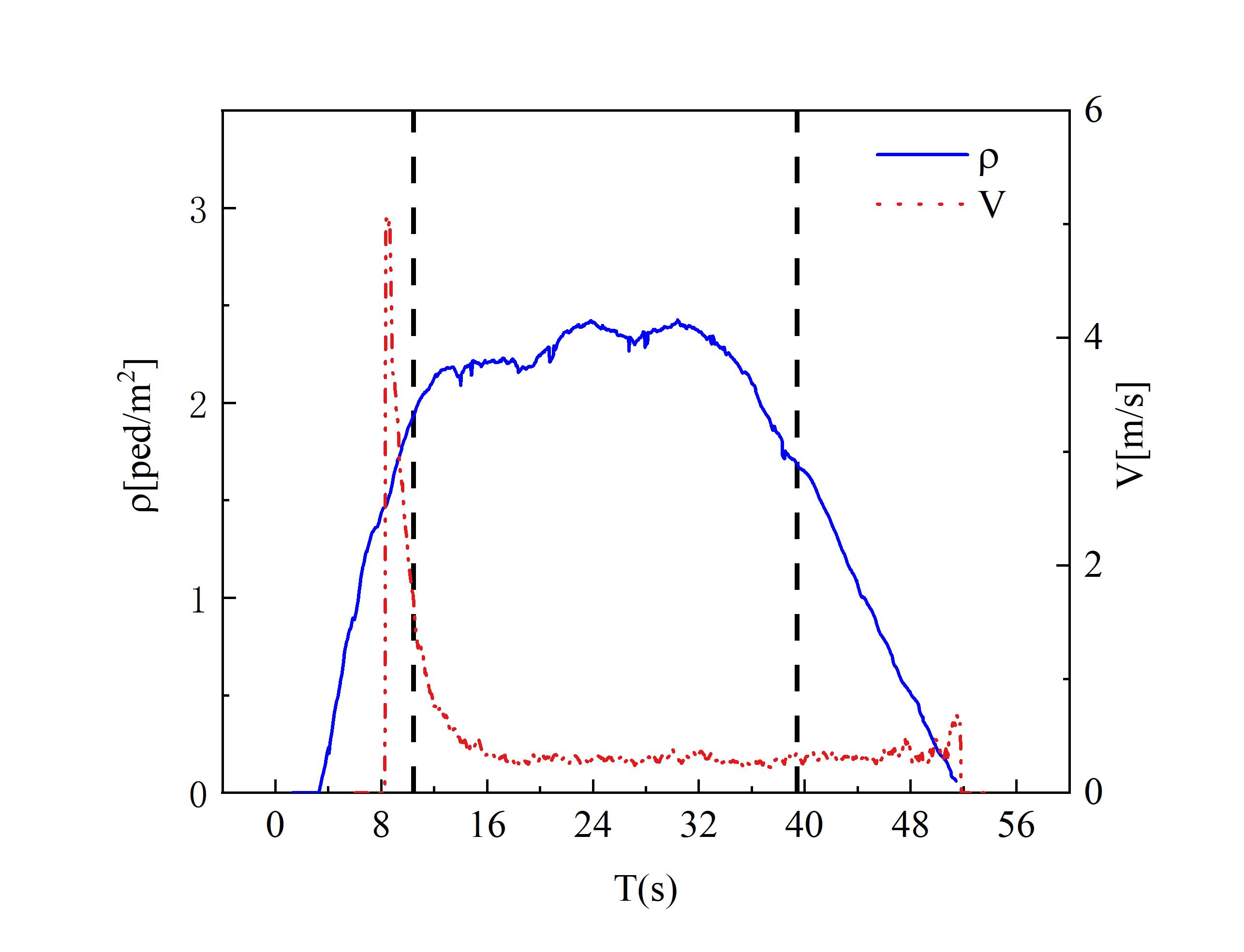}
\caption{Steady-state phase selection (no device steady-state selection process as an example; other cases are selected according to this method).}
\label{fig:Steady-state}
\end{figure}

To further quantify the effect of the diversion device on pedestrian motion characteristics, we calculate the velocity variation of pedestrians in the measurement area over time from the trajectory data and obtained the global density variation utilizing Voronoi diagrams \cite{boltes2015automatische}. Note that due to the limited number of experiments, there is a process of density map variation at the beginning and end of the experiment. This is shown in \cref{fig:Steady-state}, we refer to the selection of the steady state of the velocity density in the case of accessibility by using the modified cumulative and control chart algorithm CUSUM calculation \cite{liao2016measuring}: two vertical lines represent the beginning and end time points of the selected relative steady state, and the period between the two vertical lines is the data analysis of the selected steady state. The data analysis in \cref{subsubsec:density} takes the steady-state interval of each operating condition for calculation.

\subsubsection{Density Variation}
\label{subsubsec:density}

\begin{figure}[htbp]
\centering
\subfloat[]{\includegraphics[width=0.5\linewidth]{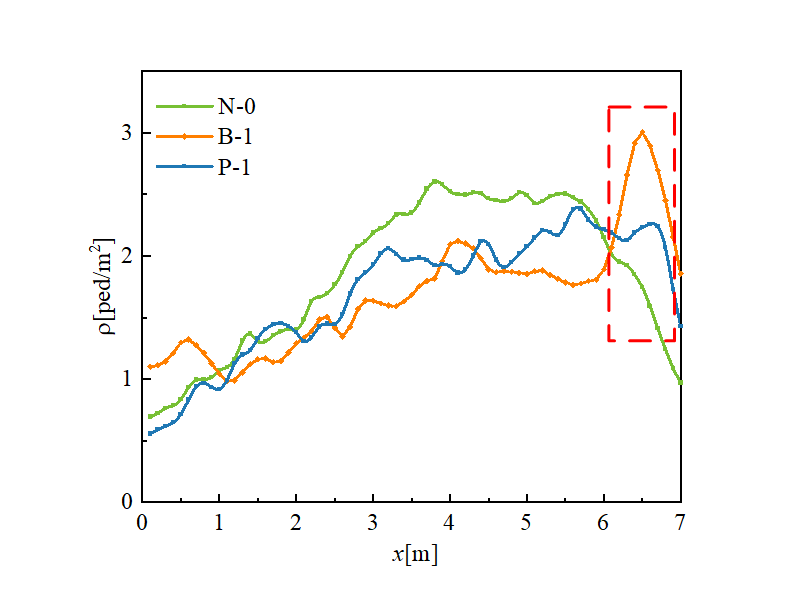}}\hfill
\subfloat[]{\includegraphics[width=0.5\linewidth]{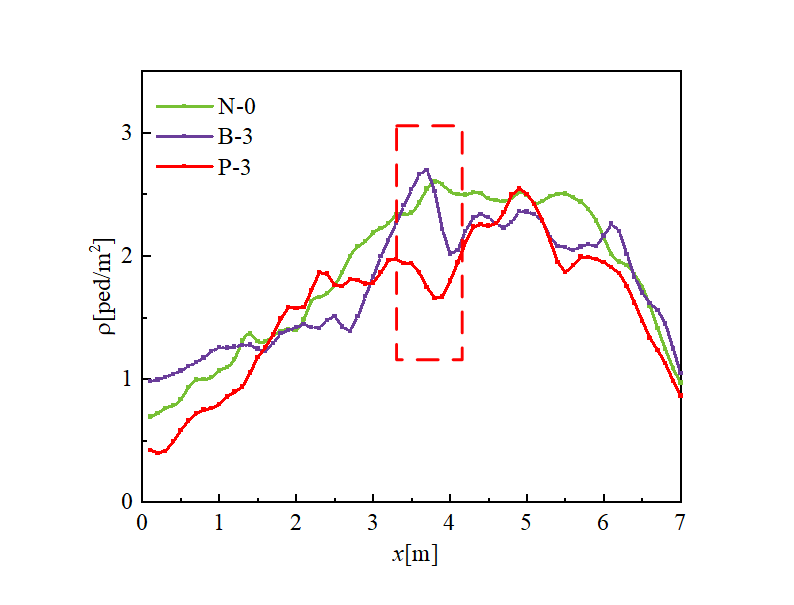}}\\
\caption{Distribution of the effect of different devices on pedestrian density (where the safety barriers are indicated by B, the shunt piles are indicated by P, and no device is indicated by N; the number represents the distance to the exit; the black box represents the peak population density of the safety barriers).}
\label{fig:density}
\end{figure}

\cref{fig:density} plots the density variation curves over time for different device settings. It can be seen that the pedestrian density at $4\sim6~m$ has been maintained at a high level without barriers of approximately $2.45~p/m^2$. The overall average density decreases after the installation of safety barriers or shunt stakes, approximately $2.04~p/m^2$ and $2.13~p/m^2$, respectively. In particular, we can see a clear peak at the end of the safety barriers when the safety barriers are installed. Comparing \cref{fig:density} (a) and \cref{fig:density} (b) and the corresponding evacuation time, we find that if the safety barriers (B-1) are set at a place closer to the exit, it tends to cause the crowd density to be too high and gather at the exit location, thus reducing the evacuation efficiency, compared to setting it at a place farther away from the exit, which can effectively alleviate this phenomenon. In the case of setting up shunt piles, the density fluctuation after leaving the barrier is less obvious and the overall density is reduced, which indicates that setting up shunt stakes has a positive effect on the enhancement of the evacuation process.

\subsubsection{Walk-stop Ratio}
As the density before the exit increases, pedestrians appear to switch between waiting and walking alternatively, which is called the walk-stop phenomenon. Helbing \etal \cite{helbing2007dynamics} analyzes videos of the pilgrimage stampede in Mecca and found that the flow of pedestrians developed from laminar to "stop-and-go", eventually forming turbulence. During the turbulence phase, the crowd gradually loses control, pedestrians are wrapped up in the crowd and move back and forth, and stampede accidents occur. Stop-and-go behavior is an intermediate stage in the development of pedestrian flow from laminar flow to turbulence, which may be a sign of the danger of pedestrian flow. Thus, we have counted the stop-and-go phenomenon in various situations, and quantitatively expressed the stop-and-go behavior of pedestrians during the movement with the waiting ratio. The waiting ratio is calculated as shown in \cref{eqn:walk_stop}.

\begin{equation}
R_w = \frac{\sum_{i=1}^N t_{wi}}{\sum_{i=1}^N T_{mi}}
\label{eqn:walk_stop}
\end{equation}

where $t_{wi}$ is the waiting time of each pedestrian, $T_{mi}$ is the movement time of each person, the numerator represents the sum of the waiting time of all pedestrians in the same working condition, and the denominator represents the sum of the walking time of all pedestrians in the same working condition. Combined with the method used by \cite{zeng2019experimental, cao2016pedestrian, ma2021spontaneous} to calculate the walk-stop ratio, we select the duration of each pedestrian at a speed below the critical speed state as the waiting time. Referring to other studies in \cite{cao2016pedestrian} that the pedestrian is stopping when the speed is below $0.1~m/s$, the critical speed is set to $0.1~m/s$ in this paper.

\begin{figure}[htbp]
\centering
\includegraphics[width=0.5\linewidth]{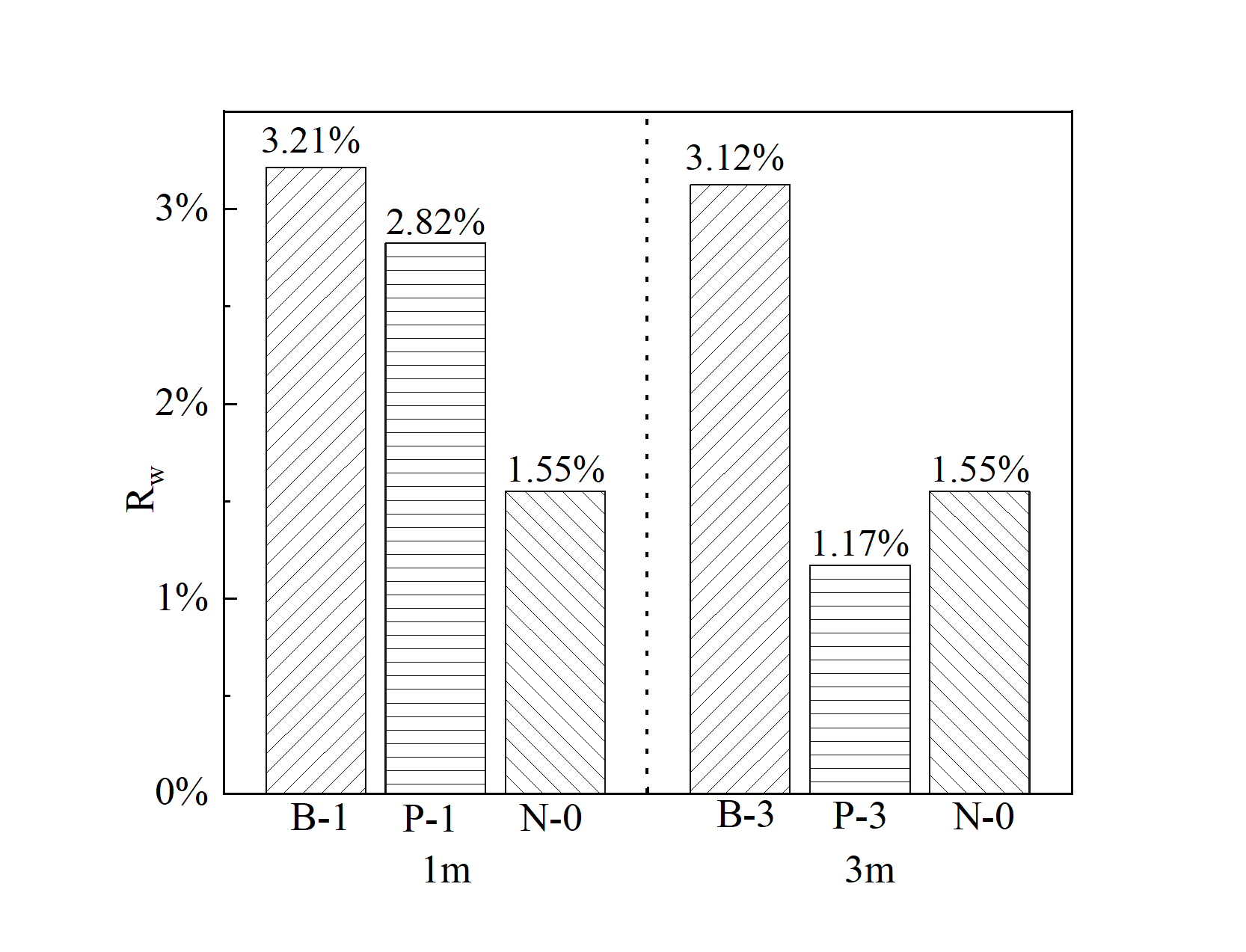}
\caption{Pedestrian waiting ratio for different conditions (the left half is $1~m$ from the exit for the device, with the figure being $3~m$ from the exit for the device; left slash represents the safety barriers, the horizontal line represents the shunt piles, right slash represents no device).}
\label{fig:waiting_ratio}
\end{figure}

As shown in \cref{fig:waiting_ratio}, when the device is set at the same location, the effect on pedestrian movement is different. When set at $1~m$ from the exit, the waiting ratio is higher than without the device because the density of pedestrians in front of the exit increases in the case of the device, and more pedestrians wait, resulting in more walk-stop phenomenon; thus, the waiting ratio increases. When the device is moved back to $3~m$ from the exit, the waiting ratio is lower than when the device is set at $1~m$. These results show that in the process of moving towards the exit, reasonable setting of the device position can make the crowd waiting for phenomenon less, the movement is more smooth, and the shunt piles can reduce the waiting ratio before the exit.

\subsection{Congestion Level}
\label{subsection:CL}

Due to the limited exit width, most pedestrians in all conditions will stall in front of the exit, and excessive crowd density will cause congestion. Evacuation behavior is limited if measured only by density, so we introduce a method for calculating the congestion level \cite{feliciani2018measurement}, which indicates the degree of pedestrian congestion by introducing rotation as an important pedestrian movement mechanism \cite{helbing2005self, feliciani2016empirical}. Following \cite{feliciani2018measurement}, we first divide the measurement area into several small grids of $0.1~m \times 0.1~m$, and each grid calculates the velocity vector to obtain a velocity vector field. Then using \cref{eqn:velocity}, the rotational value of this vector field can be calculated, and the congestion level (Cl) is the difference between the extreme values of rotation in the region divided by the average absolute velocity of the region, as follows:

\begin{equation}
\vec{R}(x,y) = (r_x, r_y, r_z) = \nabla \times \vec{v}(x, y)
\label{eqn:velocity}
\end{equation}

\begin{equation}
Cl = \frac{max(r_z)-min(r_z)}{|\vec{v}|}
\label{eqn:level}
\end{equation}

\begin{figure}[htbp]
\centering
\subfloat[$1~m$]{\includegraphics[width=0.5\linewidth]{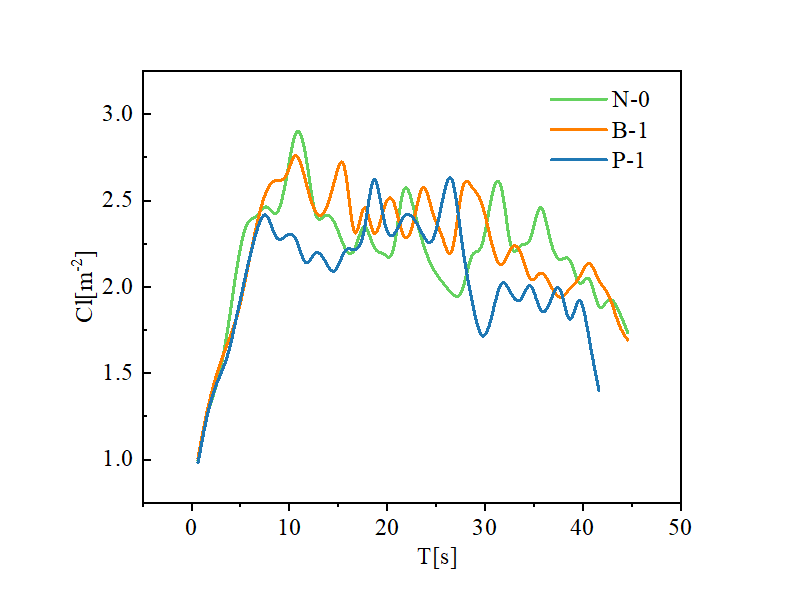}}\hfill
\subfloat[$3~m$]{\includegraphics[width=0.5\linewidth]{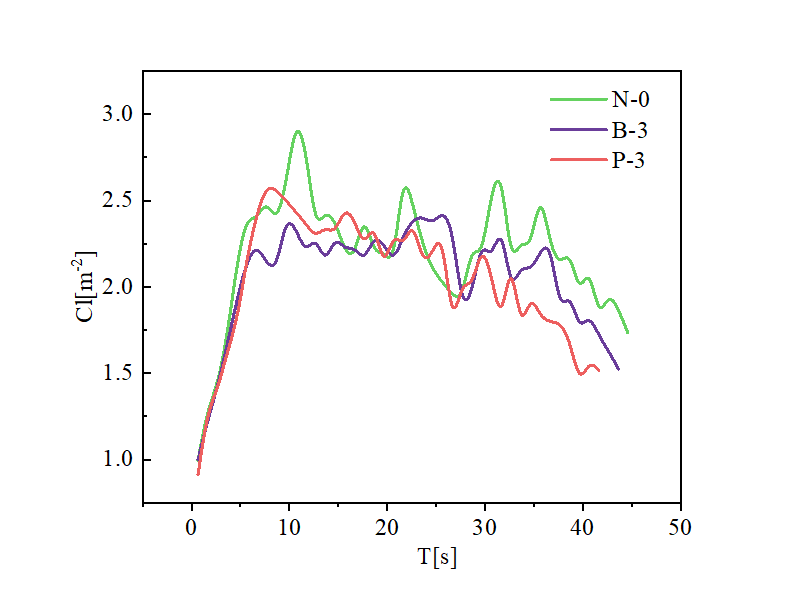}}
\caption{Variation of the congestion level at different exit distance.}
\label{fig:congestion}
\end{figure}

Defining $x$ and $y$ as the two axes of the grid being analyzed, it is possible to compute the rotational of this vector field  $|\vec{v}|$, $r_x$ and $r_y$ will be zero since $x-$ and $y-$ velocities lie on the same plane. As a consequence, only $r_z$ takes values different from zero and it will be used for the forthcoming analyses. where $max(r_z)$ and $min(r_z)$ are the maximum and minimum values of $r_z$ over a defined region and  $|\vec{v}|$ is the average absolute velocity in that region. According to \cref{eqn:velocity} and \cref{eqn:level}, we can plot the congestion level with time, as shown in \cref{fig:congestion}.

As seen from the \cref{fig:congestion}, the congestion level starts to increase with time and after reaching the peak, the fluctuations begin to decline after a period of time. The \cref{eqn:level} shows that the congestion level is mainly determined by the pedestrian spin and the average speed. After around $10~s$, the pedestrians in the measurement area have reached a steady state, and the average speed does not change much, so it is mainly the pedestrian spin that plays a decisive factor. It is obvious from \cref{fig:congestion} (b) that the congestion level without devices fluctuates more, and there is already a decreasing trend in approximately $30~s$ with devices, when pedestrians are mostly in the device area, forming a queue before the exit, and the distribution is relatively neat, while pedestrians without devices are scattered in front of the exit, and the speed direction changes more, so the congestion level still shows an increasing trend.

\begin{figure}[htbp]
\centering
\subfloat[Early stage of  experiment]{\includegraphics[width=0.5\linewidth]{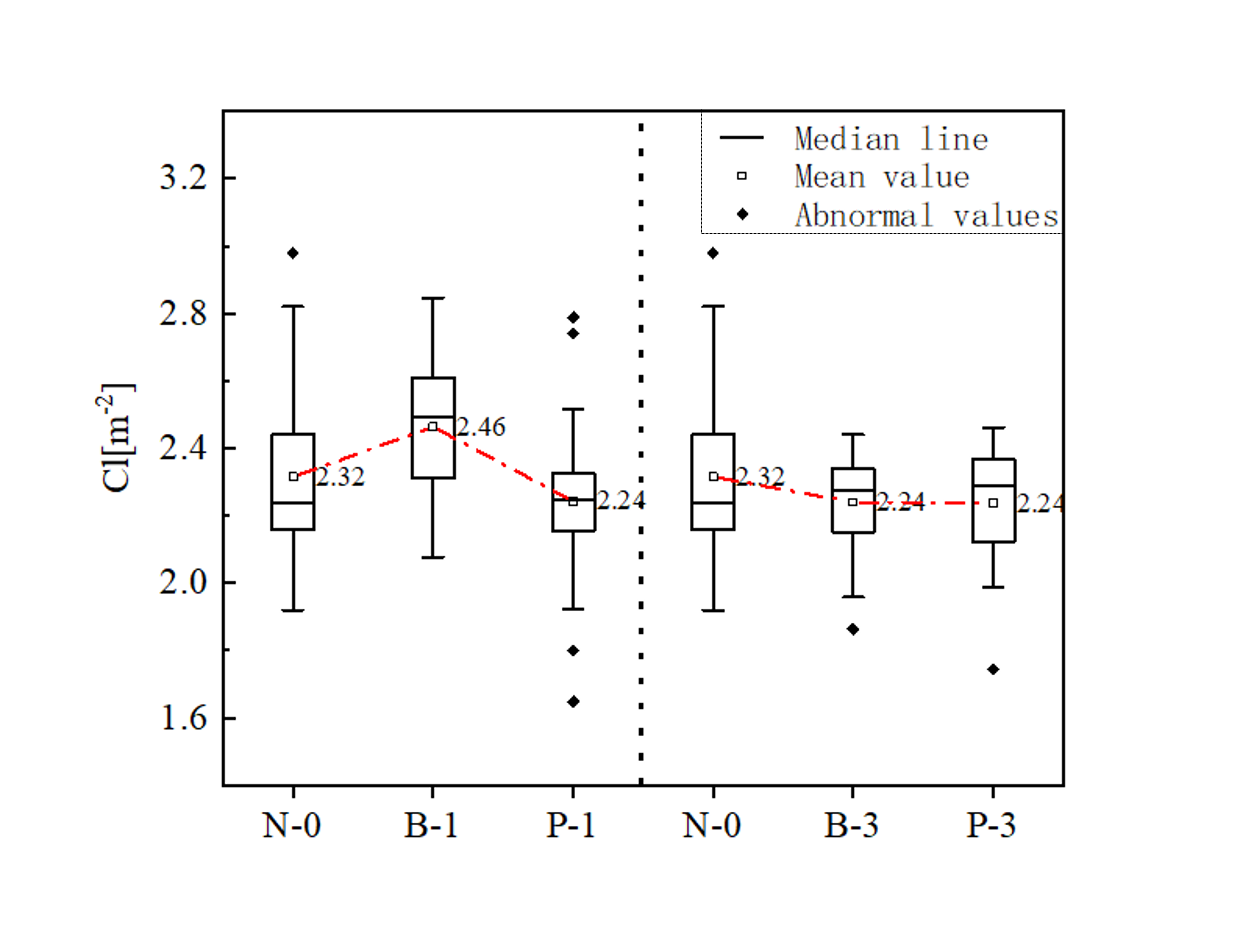}}\hfill
\subfloat[Late stage of experiment]{\includegraphics[width=0.5\linewidth]{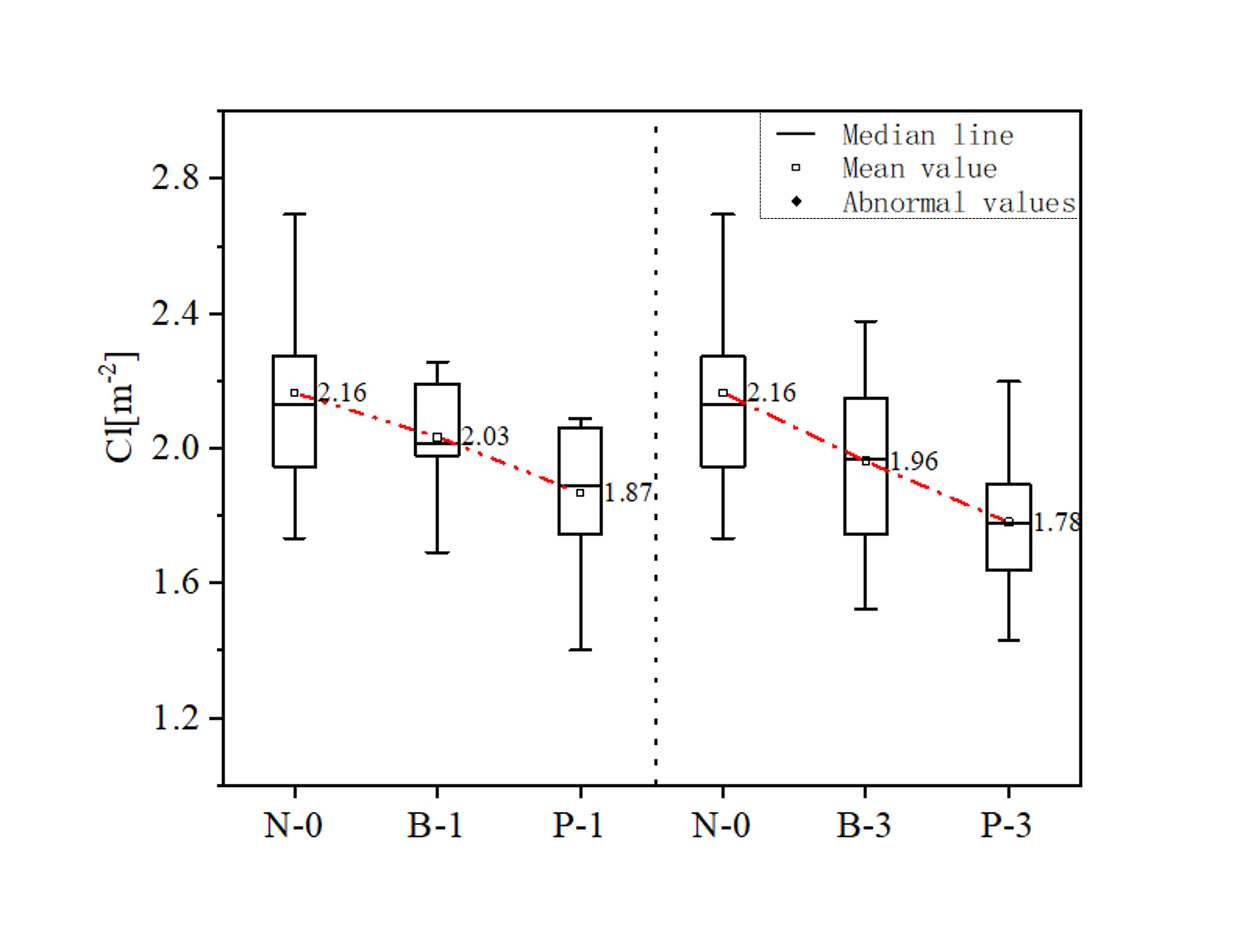}}
\caption{Congestion level of different devices before/after the experiment (the left part is the comparison diagram between the  without device and the devices placed $1~m$ away from the exit, and the right part is the comparison diagram between the without the device and the devices placed $3~m$ away).}
\label{fig:congestion_compare}
\end{figure}

It can be further observed from the above the \cref{fig:congestion} that the congestion level is in an upward trend in the early stage and enters a stable fluctuation stage after reaching the peak. Therefore, we calculate the slope of each point on the curve, find the minimum value and continue for some time. Then, this period is the congestion level decline stage, and after this period, it is called the evacuation late stage. Take $10.65~s$ when the congestion level begins to stabilize as the start time of the early stage and draw the boxplots before and after the congestion level in \cref{fig:congestion_compare}. From the box line plot, we can observe the quartiles, whiskers, and outliers of the congestion level data. From the bottom to the top of the vertical box line plot, each horizontal line represents the lower whisker, lower quartile (Q1, 25), median (50), upper quartile (Q3, 75), upper whisker, and mean (marked by black dots in the box), respectively.

In \cref{fig:congestion_compare} We can find that working conditions in the early stage of the difference are less obvious, but in the latter, can see the congestion level of each group has significant differences. In particular, in the case of setting diversion devices, the congestion level is lower than the condition of no device, and the diversion devices increase with the distance to the export decline (in diversion devices, the congestion level of the safety barriers and shunt piles decreases by $6.94\%$ and $13.43\%$, respectively). When set to $3~m$, the congestion level of the safety barriers and shunt piles decreases by $9.26\%$ and $17.59\%$, respectively. Therefore, it can be concluded that setting the diversion devices at a specific distance before the outlet can effectively suppress the outlet congestion level, and the shunt piles have a stronger inhibitory effect on the congestion level. There may be an optimal distance to make its inhibition more pronounced, which we will continue to explore in future experiments. It can be seen from the conclusion in \cref{subsec:interval} that the velocity tends to be consistent, which is conducive to personnel evacuation, and the congestion level, as an index to measure velocity fluctuation, can well reflect the degree of velocity consistency \cite{garcimartin2015flow}. It can be proven that the placement of shunt piles leads to the reduction of the degree of chaos in the later stage of evacuation: The shunt piles set at a specific distance before the exit can effectively guide pedestrians to move in a favorable direction to improve the efficiency of pedestrian evacuation.

\section{Conclusion}
\label{sec:conclusion}

In this work, we study the characteristics of pedestrian evacuation in emergencies and the differences between pedestrian evacuation by different diversion devices through experiments with diversion devices placed in front of a narrow exit. By studying the influence of the devices on the motion characteristics of pedestrians in terms of density, speed, congestion, and direction of movement, the main conclusions are as follows:

\begin{itemize}
    \item As the distance between the devices and the exit increases, the overall evacuation time shows a decreasing trend. The experimental results show that the devices can reduce the congestion level of the late evacuation by $17.59\%$, improve the order degree of pedestrians, shorten the overall evacuation time by $14.96\%$, and improve the evacuation efficiency under the best condition of the experiment..
    \item The experimental data show that the devices at $3~m$ can prompt the pedestrian to adjust the direction of movement in advance and shrink toward the exit than the devices at $1~m$, and the position of the pedestrian on both sides is different. Shunt piles can reduce the proportion of the walk-stop phenomenon. In the best case, the time interval between consecutive pedestrians in front of the exit is reduced by $17.24\%$, making the evacuation process smoother.
    \item From the results, we can see that the average density varies throughout the space, with roughly the same trend in different scenarios. The peak density of the devices at $1~m$ from the exit appears in front of the exit. The back of the devices can drive the density peak in front of the exit into a rapid decline in density.
\end{itemize}

By comparing the two kinds of diversion devices, it is found that the shunt piles are better than the safety barriers in all aspects. In actual life, shunt piles occupy a small area, are easy to move, and have the least negative impact on pedestrians while improving evacuation efficiency, so they can be applied in mass activities.

The analysis in this paper enriches the empirical data on pedestrian evacuation at bottlenecks and demonstrates that the type and distance of pre-exit devices can have different effects on pedestrian evacuation. However, the above conclusions are only based on experimental devices at two different distances. In future work, we will continue to refine the effects of devices on pedestrian evacuation at multiple distances and strengthen theoretical and modeling aspects to draw more comprehensive conclusions.

\section*{Acknowledgments}

This study was supported by the National Natural Science Foundation of China (52074252), Key R \& D Program of Anhui (202004a07020052), and Fundamental Research Funds for the Central Universities (Grant No. WK2320000050).

\bibliographystyle{unsrt}  
\bibliography{main}  

\end{document}